\documentclass[preprint]{JHEP3} % 10pt is ignored!
\usepackage{epsfig}
\usepackage{amsfonts}
\usepackage{amssymb,amsmath}
\usepackage{dsfont}
\newcommand{\Tr}[1]{\:{\rm Tr}\,#1}

\newcommand{\R}{{\ell}}

\newcommand{\mbf}[1]{{\boldsymbol {#1} }}
\newcommand{\complex}{{\mathds C}} %% complex numbers
\newcommand{\zed}{{\mathds Z}} %% integers
\newcommand{\nat}{{\mathds N}} %% natural numbers
\newcommand{\real}{{\mathds R}} %% real numbers
\newcommand{\torus}{{\mathds T}}
\newcommand{\sphere}{{\mathds S}}
\newcommand{\disk}{{\mathds D}}

\def\e{{\,\rm e}\,}

\newcommand{\id}{{1\!\!1}}

\def\ii{{\,{\rm i}\,}}
\def\dd{{\rm d}}
\def\DD{{\rm D}}

\def\beq{\begin{equation}}
\def\bee{\begin{equation}}
\def\eeq{\end{equation}}
\def\bea{\begin{eqnarray}}
\def\eea{\end{eqnarray}}
\def\bd{\begin{displaymath}}
\def\ed{\end{displaymath}}

\newcommand{\Cint}{\int\kern-10.5pt-\kern7pt}

\newcommand{\PP}{{\mathds{P}}}

\newcommand{\be}{\begin{equation}}
\newcommand{\ee}{\end{equation}}

\newcommand\fverb{\setbox\pippobox=\hbox\bgroup\verb}
\newcommand\fverbdo{\egroup\medskip\noindent%
                        \fbox{\unhbox\pippobox}\ }
\newcommand\fverbit{\egroup\item[\fbox{\unhbox\pippobox}]}
\newbox\pippobox

\title{Black Holes, Instanton Counting on Toric Singularities and
  q-Deformed Two-Dimensional Yang-Mills Theory}
\author{Luca Griguolo$^{(a)}$, Domenico~Seminara$^{(b)}$,
  Richard~J.~Szabo$^{(c)}$ and Alessandro Tanzini$^{(d)}$ \\
$^{(a)}$  Dipartimento di  Fisica, Universit\`a  di Parma and
INFN Gruppo Collegato di Parma\\
Parco Area delle Scienze 7/A, 43100 Parma, Italy\\
$^{(b)}$ Dipartimento di Fisica, Polo Scientifico Universit\`a di
Firenze and\\ INFN Sezione di Firenze\\
Via  G. Sansone 1, 50019 Sesto Fiorentino, Italy\\
$^{(c)}$ Department of Mathematics, Heriot-Watt University and\\
Maxwell Institute for Mathematical Sciences\\
Colin Maclaurin Building, Riccarton, Edinburgh EH14 4AS, UK\\
$^{(d)}$ Scuola Internazionale Superiore di Studi Avanzati and
INFN Sezione di Trieste\\ Via Beirut 4, 34014 Trieste, Italy \\
\email{griguolo@fis.unipr.it , seminara@fi.infn.it ,
  R.J.Szabo@ma.hw.ac.uk , tanzini@sissa.it }}
\received{\today}               %%

\accepted{\today}               %% These are for published papers.
\preprint{ {\tt HWM-06-37} \ \ {\tt EMPG-06-08} \ \ {\tt
    SISSA-61/2006/FM}
\\ \hepth{0610155}} % OR: \preprint{Aaaa/Mm/Yy\\Aaa-aa/Nnnnnn}
\date{data}
\abstract{We study the relationship between instanton counting in
$\mathcal{N}=4$ Yang-Mills theory on a generic four-dimensional toric
orbifold and the semi-classical expansion of q-deformed Yang-Mills
theory on the blowups of the minimal resolution of the orbifold
singularity, with an eye to clarifying the recent proposal of using
two-dimensional gauge theories to count microstates of black holes in
four dimensions. We describe explicitly the instanton contributions to
the counting of D-brane bound states which are captured by the
two-dimensional gauge theory. We derive an intimate relationship
between the two-dimensional Yang-Mills theory and Chern-Simons theory
on generic Lens spaces, and use it to show that the correct instanton
counting is only reproduced when the Chern-Simons contributions are
treated as non-dynamical boundary conditions in the D4-brane gauge
theory. We also use this correspondence to discuss the counting of
instantons on higher genus ruled Riemann surfaces.}


\begin{document}

\section{Introduction}

A renewed impetus into the description of BPS black hole microstates in
four dimensions has been sparked by the OSV
conjecture~\cite{Ooguri:2004zv} which equates black hole entropy in
Type~IIA string theory compactified on a Calabi-Yau threefold $\cal M$
to the modulus squared of the topological string partition function on
$\cal M$. The black hole is constructed by wrapping D2-branes around
arbitrary two-cycles of $\cal M$ and D4-branes around a four-cycle
which is a fixed ample divisor of $\cal M$. With respect to a fixed
basis of two-cycles in $H_2({\cal M},\zed)$, and a dual basis of
four-cycles in $H_4({\cal M},\zed)$, the D2 and D4~branes carry
electric and magnetic charges $\mbf Q_2,\mbf Q_4\in\zed^n$ where
$n=h^{1,1}({\cal M})$. We also specify the D0-brane charge
$Q_0\in\zed$ and turn off the D6-brane charge. The black hole
partition function then takes the symbolic form
\beq
Z_{\rm BH}(\mbf Q_4,\mbf\varphi_2,\varphi_0)=\sum_{\mbf
  Q_2\in\nat_0^n}~\sum_{Q_0\in\nat_0}\,
\Omega(\mbf Q_4,\mbf Q_2,Q_0)~\e^{-Q_0\,\varphi_0-\mbf Q_2\cdot
\mbf\varphi_2}
\label{ZBHsymb}\eeq
where $\Omega(\mbf Q_4,\mbf Q_2,Q_0)$ is the indexed degeneracy of BPS
states in spacetime with the specified charges, and
$\varphi_0,\mbf\varphi_2$ are chemical potentials. The OSV conjecture
then equates (\ref{ZBHsymb}) for large black hole charges to the
topological string amplitude $|Z_{\rm top}(\mbf t,g_s)|^2$, where
$\mbf t\in\complex^n$ are the K\"ahler parameters of $\cal M$ and the
various moduli between the two partition functions are related by the
attractor mechanism.

The index $\Omega(\mbf Q_4,\mbf Q_2,Q_0)$ can be computed by counting
BPS states in the supersymmetric gauge theory on the
D4-branes~\cite{Vafa:1995bm}, where the D0-branes are interpreted as
instantons and the D2-branes as sources of magnetic flux endowing the
instantons with non-trivial first Chern class. Then (\ref{ZBHsymb})
coincides with the partition function of ${\cal N}=4$ topological
Yang-Mills theory in four-dimensions, summed over all topological
sectors and with the insertion of observables giving the inclusion of
D0 and D2~brane charges. The structure of this gauge theory partition
function has been recently argued
in~\cite{Vafa:2004qa,Aganagic:2004js} to simplify drastically in the
case of a local Calabi-Yau space which is a rank~$2$ normal bundle
over a compact Riemann surface $\Sigma$, and the D4-brane worldvolume
is given by the total space of a non-trivial holomorphic line bundle
over $\Sigma$. It is argued that the partition function
(\ref{ZBHsymb}) localizes onto field configurations which are
invariant under the natural $U(1)$ action on the fibres of this line
bundle, and the four-dimensional gauge theory reduces to a
two-dimensional gauge theory on the base $\Sigma$ called q-deformed
Yang-Mills theory~\cite{Boulatov}--\cite{Klimcik:1999kg}. Various
aspects of the black hole partition function on toric Calabi-Yau
threefolds from this remarkable
two-dimensional point of view are analysed
in~\cite{Vafa:2004qa,Aganagic:2004js,Aganagic2:2005,Jafferis:2006ny}.

In this paper we will analyse in detail the problem of computing the
black hole partition function (\ref{ZBHsymb}) using the sewing
formalism of q-deformed Yang-Mills theory for the most general toric
singularity $X(p,q)$ in four dimensions.\footnote{A word of caution about
  notation. In the literature on q-deformed gauge theory the symbol
  $q$ is used to denote the q-deformation, which in the topological
  string setting is given by $q=\e^{-g_s}$. In this paper we will only
  use $q$ to denote the integer modulus of the toric four-manifold,
  always writing the q-deformation explicitly as $\e^{-g_s}$ with the
  usual identification $g_s=g_{\rm YM}^2/2$ between the string and
  four-dimensional Yang-Mills coupling constants. Accordingly, we also
  avoid using the standard notation $q$ for the arguments of modular
  forms which typically arise in four-dimensional instanton calculations.} This
construction extends the $A_{k}$ ALE spaces which were considered
in~\cite{Aganagic2:2005}. It also includes the four-manifolds $X(p,1)$
which are the total spaces of the holomorphic line bundles ${\cal
  O}_{\PP^1}(-p)$ and for which the relevant two-dimensional gauge
theory is q-deformed Yang-Mills theory on the sphere which was studied
in great detail
in~\cite{Aganagic:2004js,Arsiwalla:2005jb}--\cite{Caporaso:2005np}.
Our results are in agreement with the recent analysis
in~\cite{fcr} of the black hole partition function (\ref{ZBHsymb})
using direct instanton calculations in the four-dimensional gauge
theory.

One of our main computations is the modular inversion of the heat
kernel representation of the q-deformed partition function which casts
it as a sum over two-dimensional Yang-Mills instantons living on the
blowups of the minimal resolution of the toric singularity. This
resummation is necessary to match the topological expansion
(\ref{ZBHsymb}), and we immediately find problems with the black hole
interpretation of the two-dimensional gauge theory. The
semi-classical expansion of the two-dimensional gauge theory contains
terms which cannot simply correspond to an indexed degeneracy
$\Omega$. We identify part of the expansion with the value of the
Chern-Simons partition function on the boundary of the non-compact
space $X(p,q)$, which is a generic three-dimensional Lens space
$L(p,q)$. To match with the four-dimensional instanton computation we
must follow the standard prescription of summing over the admissible
{\it non-dynamical} boundary conditions on the gauge fields, whose
asymptotic values are governed by Chern-Simons gauge theory. This
amounts to identifying that part of the two-dimensional amplitude
which corresponds to the perturbative expansion of the Chern-Simons
theory about a given vacuum. We will find that, when these terms are
stripped and only the classical Chern-Simons contributions are
retained, the q-deformed gauge theory reproduces {\it exactly} the
contributions given in~\cite{fcr} from ``fractional'' instantons which
are stuck at the singularity of $X(p,q)$. In particular, it is not
entirely clear exactly how the two-dimensional formalism can reproduce
the remaining contributions, such as those coming from instantons
which are free to propagate throughout the four-dimensional space.

In the course of this analysis we are faced with the derivation of the
nonabelian localization formula for the Chern-Simons partition
function on a generic Lens space $L(p,q)$ (we have not found a
complete and general calculation in the literature). We carry out this
computation in detail by using Seifert fibration techniques to
evaluate the classical contributions and surgery methods to compute
the fluctuation determinants. In particular, we derive the explicit
mapping between Chern-Simons vacua and two-dimensional Yang-Mills
connections, and hence with fractional instantons in four
dimensions. We also briefly examine the problem of counting instantons
on ruled Riemann surfaces for genus $g\geq1$, which are non-toric
four-manifolds for which little is known about the structure of
instantons. We use the prescription given above to predict the
structure of the full $U(1)$ partition function in four dimensions for
any genus, and to predict the contributions from fractional instantons
in the nonabelian case for genus $g=1$. In particular, we conclude
that the $U(N)$ partition function does not seem to factorize into a
$U(1)^N$ contribution, as it does in the genus~$0$ cases.

The organisation of this paper is as follows. In
Section~\ref{InstSect} we review the structure of four-dimensional
instanton partition functions on the toric manifolds $X(p,q)$ and in
particular some of the results of~\cite{fcr}. In
Section~\ref{qYMToric} we construct the pertinent q-deformed gauge
theory amplitude and describe how to extract the four-dimensional
instanton contributions. In Section~\ref{CSLpq} we work out the
semi-classical expansion of Chern-Simons gauge theory on the Lens
spaces $L(p,q)$. In Section~\ref{HigherGenus} describe some
analogous computations on the ruled Riemann surfaces. In
Section~\ref{Conclusions} we summarize our findings. Finally, some
technical details of our calculations are summarized in an appendix at
the end of the paper.

\section{D-Brane Partition Function on Toric Orbifolds\label{InstSect}}

In this section we will study the partition function of a bound system
of D0--D2--D4~branes where the D4-branes wrap a four-cycle $X$
of a local Calabi-Yau threefold given by the total space of the
canonical line bundle $K_X$. We take $X$ to be a smooth four-dimensional
manifold given by the minimal resolution of the quotient space
$\mathds{C}^2/\Gamma$, where $\Gamma\cong\zed_p$ is a generic finite
cyclic group. The action of $\Gamma$ on the coordinates $(z,w)$ of
$\complex^2$ can be linearized locally as \beq \big(z\,,\,w\big)
~\longmapsto~
\big(\e^{2\pi \ii q/p}\,z\,,\,\e^{2\pi \ii /p}\,w\big)
\label{gamma} \eeq where
$(p,q)$ are coprime integers with $p>q>0$. The orbifold action
(\ref{gamma}) generates an $A_{p,q}$ singularity at the origin of
$\complex^2$, whose minimal resolution (known as the Hirzebruch-Jung
resolution) gives rise to a smooth four-dimensional manifold $X(p,q)$
called a Hirzebruch-Jung space~\cite{BPVdeV}.

This space contains a chain of $\ell$ exceptional divisors at the
origin given by projective lines $\mathds{P}^1$ whose intersection
numbers are summarized by the (generalized) Cartan matrix \beq
C=-\begin{pmatrix} - e_1 & 1 & 0 & \cdots &0\\
1 & -e_2  & 1& \cdots &0\\
0 & 1 & - e_3 &\cdots&0\\
\vdots &\vdots & \vdots &\ddots&\vdots\\0&0&0&\cdots&-e_\ell
\end{pmatrix} \ .
\label{InterMatrix}\eeq  The moduli of the self-intersection numbers
$e_i\ge 2$, $i=1,\ldots,\ell$ are obtained by expanding the rational
number $\frac pq>1$ in a simple continued fraction
\beq \frac pq=[e_1,\dots,e_\ell]:=
e_1-{1\over\displaystyle e_{2}- {\strut
1\over \displaystyle e_{3}- {\strut 1\over\displaystyle\ddots {}~
e_{\ell-1}-{\strut 1\over e_\ell}}}} \label{pqcontfrac} \eeq with
$e_1$ the smallest integer $>\frac pq$, and so on. For
example, for $q=1$ there is only one exceptional divisor with
self-intersection number $-e_1=-p$ and the manifold $X(p,1)$ can be
regarded as the total space of a holomorphic line bundle ${\cal
  O}_{\PP^1}(-p)$ of degree $p$ over $\mathds{P}^1$. The other
limiting case $q=p-1$ corresponds to an $A_{p-1}$ ALE space, which
contains a chain of $\ell=(p-1)$ $\mathds{P}^1$'s each with
self-intersection number $-e_i=-2$, and in this case (\ref{InterMatrix})
coincides with the Cartan matrix of the $A_{p-1}$ Dynkin diagram.

Standard arguments~\cite{Bershadsky:1995qy} show that the gauge
theory living on $N$ D4-branes wrapping $X(p,q)$ is a $U(N)$ Vafa-Witten
topologically twisted ${\cal N}=4$ Yang-Mills theory~\cite{Vafa:1994tf}.
In this context, the D0-branes are interpreted as instantons of
the four-dimensional gauge theory. These instantons can also have
a non-vanishing first Chern class due to the presence of D2-branes
wrapping the exceptional divisors which generate a non-trivial
magnetic flux on the D4-brane worldvolume. Under suitable
assumptions~\cite{Vafa:1994tf} the twisted ${\cal N}=4$ Yang-Mills
partition function computes the Euler number of the instanton moduli
space. The powerful toric localization techniques developed in recent
years~\cite{Nekrasov:2002qd}--\cite{Fucito:2005wc} have enabled the
computation of this partition function for ALE
spaces~\cite{Fucito:2004ry} and for the total spaces of the ${\cal
  O}_{\mathds{P}^1}(-p)$ bundles with
$p=1,2$~\cite{Nakajima:2003pg,Sasaki:2006vq}. However, for generic
$A_{p,q}$ singularities an explicit description of the instanton
moduli space is not available at the moment and the direct evaluation
of the complete instanton partition function on Hirzebruch-Jung spaces
is still an open problem.

In this paper we will address this problem from a somewhat
different perspective. According to the proposal
of~\cite{Aganagic:2004js,Aganagic2:2005}, one can localize the
four-dimensional path integral via the natural $U(1)$ action on the
fibres of the normal bundles over the $\PP^1$'s.
In this way one reduces the D4-brane gauge theory to a q-deformed
Yang-Mills theory living on the exceptional divisors which arise from the
minimal resolution of the toric orbifold singularity. As we will see
in the following, the two-dimensional computation gives results in
agreement with the direct instanton counting presented recently
in~\cite{fcr}, where the instanton partition functions on $A_{p,q}$
toric orbifolds are described by assuming some factorization
properties which we review below.

For ALE spaces an explicit description of the instanton moduli
space in terms of ADHM data has been derived in~\cite{kronaka} and
reinterpreted in terms of D-brane bound states
in~\cite{Douglas:1996sw}. Let us recall some features of this
construction which will be useful in the following. The first and
second Chern characters of a $U(N)$ instanton gauge bundle ${\cal E}$
over an ALE space $X_p:=X(p,p-1)$ are given by \bea
{\rm ch}_1 ({\cal E}) &=& \sum_{i=0}^{p-1}\, u_i~{\rm ch}_1 ({\cal
  T}^i)  \ , \nonumber \\[4pt]
{\rm ch}_2 ({\cal E}) &=& \sum_{i=0}^{p-1}\, u_i~{\rm ch}_2 ({\cal
T}^i) - \frac{K}{p}\, \Omega_{X_p} \qquad \mbox{with} \quad
K = \sum_{i=0}^{p-1}\, k_i \ ,
\label{chern} \eea where ${\cal T}^i$ are principal $U(1)$-bundles of
degree $i$ corresponding to the tautological bundles associated to
each of the exceptional divisors~\cite{gocho}. In particular, ${\cal
  T}^0$ is the trivial line bundle. In eq.~(\ref{chern}), ${\rm ch}_1
=c_1$ and ${\rm ch}_2 =\frac12\,c_1^2 - c_2$, where $c_1$ and $c_2$ are the
 first and second Chern classes, respectively, and $\Omega_{X_p}$ is the
 unit volume form on $X_p$. The coefficients $u_i$, $i=0,1,\ldots, p-1$
 are given in terms of partitions $K=\sum_i\,k_i$ and $N = \sum_i\,N_i$ of
 the numbers of D0 and D4~branes, respectively, into
the $i$-th irreducible representation of $\zed_p$ as \be u_i =-
C_{ij}\, \int_{X_p}\, c_1({\cal E}) \wedge c_1({\cal T}^j) = N_i + k_{i+1} +
k_{i-1} - 2 k_i \ . \label{ui} \ee On the ALE space $X_p$ one can
distinguish between two classes of instantons, the \textit{regular}
and \textit{fractional} instantons~\cite{Fucito:2001ha}.

Regular instantons live in the regular representation
$k_0=k_1=\ldots=k_{p-1}=k$ of the orbifold group $\zed_p$.
As such, they are free to move together with their orbifold images on
the whole space $X_p$ and their moduli space for gauge group $U(1)$
coincides with the Hilbert scheme $X_p^{[K]}$ of $K= k\,p $
points on $X_p$. The Poincar\'e polynomial of this space is well-known
and is given by~\cite{nakabook}
\be P\big(t\,\big|\,X_p^{[K]}\big) = \prod_{m=1}^\infty\,
\frac{1}{\left(1-\e^{2\pi\ii m\,\tau}\,t^{2m}\right)^{p-1}\,\left(1-
\e^{2\pi\ii m\,\tau} \,t^{2m-2}\right)} \ ,
\label{PoincareALE}\ee
where $\tau = \frac{4\pi \ii}{g_{\rm YM}^2}
+ \frac{\theta}{2\pi}$ is the complexified gauge coupling. The
expression (\ref{PoincareALE}) is a function of the usual Boltzmann
weight of regular instantons in the supersymmetric Yang-Mills path
integral. By putting $t=1$ in (\ref{PoincareALE}) one gets the Euler
characteristic of the moduli space of regular $U(1)$ instantons given
by \be Z^{U(1)}_{\rm
  reg} = \frac{1}{\hat\eta(\tau)^{p}} \qquad \mbox{with} \quad
\hat\eta(\tau):= \prod_{m=1}^\infty\,\left(1-\e^{2\pi\ii m\,\tau}
\right) \ . \label{reg} \ee
The generic $U(N)$ partition function is given by the $N$-th power of
(\ref{reg}).

Fractional instantons are instead stuck at the orbifold singularity
and have no moduli associated to their position in the
four-dimensional space $X_p$. More precisely, they correspond to the
non-trivial self-dual abelian gauge connections $A^{\rm
  frac}$ of the tautological bundle with curvature \be F_A^{\rm frac}
= -2\pi \ii\,\sum_{i=0}^{p-1}\, u_i\, c_1({\cal T}^i) \
. \label{tauto} \ee From (\ref{chern}) we then immediately realize
that their contribution to the path integral is weighted in terms of
the intersection matrix \be I^{ij}:=
\int_{X_p}\, c_1({\cal T}^i) \wedge c_1({\cal T}^j)=-
\big(C^{-1}\big)^{ij} \ , \label{C-1}
\ee where $C^{-1}$ is the inverse of the Cartan matrix
(\ref{InterMatrix}). Since the four-dimensional space
$X_p$ is non-compact, the Cartan matrix is not
unimodular and so its inverse generally has rational-valued elements (see
Appendix~A). Thus fractional instantons indeed have a fractional
charge. The contribution of fractional $U(1)$ instantons to the ${\cal
  N}=4$ partition function on ALE spaces has been written elegantly
in~\cite{fujii,fcr} by rewriting the second Chern character in
(\ref{chern}) as
\beq
{\rm ch}_2({\cal E})=\sum_{i=0}^{p-1}\,\big(C^{-1}\big)^{ii}\,
u_i-\frac Kp\,\Omega_{X_p}=\mbox{$\frac12$}\,\big(C^{-1}\big)^{ij}\,
u_i\,u_j
\label{ch2id}\eeq
to get
\be Z^{U(1)}_{\rm frac} = \sum_{\mbf u\in
  \mathds{Z}^{p-1}} \,
\e^{\pi\ii \tau\, u_i\, (C^{-1})^{ij}\,u_j}~\e^{-u_i\,z^i } \ ,
\label{frac0} \ee
where $z^i = (C^{-1})^{ij}\,(\varphi_2)_j$ is the contribution of the
magnetic fluxes associated to the D2-branes with chemical potentials
$(\varphi_2)_i$. The result for general $U(N)$ gauge group can again
be obtained by simply taking the $N$-th power of
(\ref{frac0})~\cite{fujii,fcr}.

In~\cite{Fucito:2004ry} it was observed that the regular and
fractional instanton contributions factorize in the evaluation of
the ${\cal N}=4$ partition function on ALE spaces. This result has
been further developed and established on a firm mathematical basis
in~\cite{fujii}. Thus the full partition function on ALE spaces is
given simply by the $N$-th power of the product of (\ref{reg}) and
(\ref{frac0}). In~\cite{fcr} analogous formulas are proposed
for the regular and fractional instantons on more general $A_{p,q}$
toric orbifolds. The results which follow in the next section indicate
that an analogous factorization takes place as well for these more
general four-manifolds, even though a more direct analysis is required
to properly confirm this property.

\section{q-Deformed Gauge  Theory on Toric
  Singularities\label{qYMToric}}

In this section we will evaluate the q-deformed Yang-Mills
partition function living on the minimal resolution of generic
$A_{p,q}$ orbifold singularities. After carefully resolving some
subtleties, our results will correctly reproduce the contributions of
fractional instantons to the four-dimensional gauge theory partition
function of the previous section. This is what one would naturally
expect, since the fractional instantons are bounded to the exceptional
divisors of the four-dimensional geometry. Moreover, it follows from
eq.~(\ref{tauto}) that there is a one-to-one correspondence between
fractional instantons on $X(p,q)$ and classical solutions of the
q-deformed gauge theory which are obtained as configurations of
magnetic monopoles on the $\PP^1$'s (i.e. monopole connections on the
tautological line bundles $\mathcal{T}^i$). We will describe this
correspondence in more detail in Section~\ref{CSLpq}.

\subsection{Sewing Construction of the Partition Function\label{Sewing}}

We will begin by computing the partition function of ${\cal N}=4$
topologically twisted Yang-Mills theory on the Hirzebruch-Jung spaces
$X(p,q)$ following the approach proposed
in~\cite{Aganagic:2004js,Aganagic2:2005}. This method was originally
developed in~\cite{Aganagic:2004js} for the $q=1$ cases and then later
extended to the $A_k$ ALE spaces in~\cite{Aganagic2:2005}. More
general black hole microstate counting was also attempted
in~\cite{Aganagic2:2005} using the same strategy, by considering
theories derived from more complicated configurations of D4-branes on
toric Calabi-Yau threefolds.

The main idea underlying the computation consists in cutting
the four-manifold $X(p,q)$ into pieces where the theory is simple
enough to solve explicitly. Then, thanks to the topological nature
of the gauge theory, one glues the pieces back together using an appropriate
set of rules. The Hirzebruch-Jung spaces can be obtained by patching
together $\ell$ copies of $\mathds{C}^2$, suggesting that one should be
able to derive the relevant Yang-Mills amplitudes by sewing
topological amplitudes on $\mathds{C}^2$. Since both spaces $\mathds{C}^2$
and $X(p,q)$ have $\torus^2$ isometries, the four-dimensional
gauge theory path integral should localize onto fixed points of these
torus actions. Based on this observation, a simple set of local rules for
constructing four-dimensional amplitudes in terms of q-deformed
two-dimensional Yang-Mills theory was proposed
in~\cite{Aganagic2:2005}.

The important building block in the construction is the topological
amplitude on $\mathds{C}^2$. By regarding $\mathds{C}^2$ as a
$\torus^2$ fibration over $\real^2$, it can be written as \beq {\cal
Z}(U,V)=\sum_{R,Q}\,S_{R,Q}~{\rm Tr}_{R}(U)~{\rm Tr}_{Q}(V)
\eeq where $U$ and $V$ represent the holonomies of the
four-dimensional gauge field along the boundaries of the two disks
which are fixed by the torus action~\cite{Aganagic2:2005}. The sum runs
over the irreducible representations $R,Q$ of the $U(N)$ gauge group
that label the boundary conditions on the gauge field through \beq
\int_{M_1} \,F_a=\mbox{$\frac12$}\,
n_a(R)\,g_{\rm YM}^2 \qquad \mbox{and} \qquad
\int_{M_2}\, F_a=\mbox{$\frac12$}\,
n_a(Q)\,g_{\rm YM}^2 \ , \eeq where $n_a(R)$
is the length of the $a$-th row in the Young tableau of $R$
shifted by $\frac{1}{2}\,(N+1)-a$. The two-dimensional manifolds
$M_1$ and $M_2$ are respectively the ``fiber component'' and the
``base component'' of $\mathds{C}^2$ regarded as a torus bundle.
Finally, the quantity $S_{R,Q}$ is the basic correlator \beq
\label{Corr}S_{R,Q}=\Bigl\langle{\rm Tr}_{R}\,\exp\Bigl(-\ii\int_{M_1}\,
 F\Bigr)~{\rm Tr}_{Q}\,\exp\Bigl(\ii\int_{M_2}\, F\Bigr)\Bigr\rangle\eeq
 carrying the dynamical information of the topological Yang-Mills
 theory. Its explicit expression in terms of group theoretical data is
 given below.

The complete partition function on the toric four-manifold $X(p,q)$ is
now gotten by appropriately gluing the patches together. Every time
that two disks are glued together along their boundaries (with
opposite orientation) a $\mathds{P}^1$
appears, corresponding to a partial resolution of the orbifold
singularity described in the previous section. Sewing the boundary
holonomies is achieved by integrating over them as \beq
{\cal Z}_{\PP^1}\big(V\,,\,V'\,\big)=
\int_{U(N)}\,\dd U~{\cal Z}\big(V\,,\,U\big)\,
{\cal Z}\big(U^{-1}\,,\,V'\,\big)\eeq in the invariant Haar measure on
the unitary group $U(N)$, and using standard orthogonality properties
of the group characters $\Tr_R(U)$. However, the amplitude
(\ref{Corr}) is expressed using coordinates in which
$\mathds{C}^2$ is a trivial fibration over both $M_1$ and $M_2$. For
the generic Hirzebruch-Jung spaces, the normal bundle to the $i$-th
exceptional divisor is ${\cal O}_{\PP^1}(-e_i)$ corresponding to its
non-trivial self-intersection number in (\ref{InterMatrix}). It is argued
in~\cite{Aganagic:2004js} (see
also~\cite{Vafa:2004qa,Blau:2006gh}--\cite{Caporaso:2006kk})
that the dynamical effect of the non-trivial fibration ${\cal
  O}_{\PP^1}(-e_i)$ is encoded in the term $T_R^{e_i}$ which
accompanies the gluing operation creating the corresponding
$\mathds{P}^1$, where
\beq
T_R=\e^{-\frac{g_{\rm YM}^2}{4}\,C_2(R)}
\label{TRdef}\eeq
and $C_2(R)$ is the second Casimir invariant of the representation
$R$. The presence of this term can also be
interpreted~\cite{Aganagic2:2005} as an annulus insertion, within the
general framework proposed by~\cite{Bryan:2004iq} to compute the
relevant amplitudes using two-dimensional topological quantum field
theory.

The resulting partition function on $X(p,q)$ is therefore a simple
generalization of the partition function on $A_k$ ALE spaces
constructed in~\cite{Aganagic2:2005}. The difference here is that, in
generating the chain of $\mathds{P}^1$'s by gluing disks, the
self-intersection moduli $e_i$ are generically different from~$2$. At
the ``ends'' of the chain we should turn off the gauge fields by
taking trivial holonomies on the external disks, i.e. trivial
representations $R=0$. In this way the partition function on the
Hirzebruch-Jung space takes the form
\bea
Z_{U(N)}^{q{\rm YM}}\big(X(p,q)\,,\,g_{\rm YM}^2\big)&=&
\sum_{R_1,\dots,R_\ell}\,
S_{0,R_1}\,S_{R_1,R_2}\,\cdots\,S_{R_{\ell-1},R_\ell}\,S_{R_\ell,0}~
T_{R_1}^{e_1}\,\cdots\,T_{R_\ell}^{e_\ell}\nonumber\\ &&
\qquad\qquad\times~\e^{-\ii\sum_i\,\theta_i\,C_1(R_i)} \ .
\label{Zpq}\eea
In (\ref{Zpq}) we have inserted one independent two-dimensional
$\theta$-angle $\theta_i$, $i=1,\dots,\ell$ for each exceptional
divisor, owing to the fact that the divisors define independent
homology two-cycles in $H_2(X(p,q),\zed)\cong\zed^\ell$. In the
black hole context they are related to chemical potentials for
D2-branes wrapping the divisors. We will see this explicitly in
Section~\ref{2Dto4D} below, but for the moment they simply weight here
the $U(1)$ fluxes through the $\mathds{P}^1$'s represented by the
first Casimir invariant $C_1(R)=\sum_a\,n_a(R)$ of the representation
$R$~\cite{Aganagic2:2005}. Note that for $q=1$ one has $\ell=1$ and
$e_1=p$, and (\ref{Zpq}) reduces to the partition function of
q-deformed Yang-Mills theory on the sphere~\cite{Aganagic:2004js}.

It remains to write down explicit formulas for the amplitudes
(\ref{Corr}) and (\ref{TRdef}) above. Let $\hat n_a(R)$ be the weight
vector classifying an irreducible representation $R$ of the gauge
group $U(N)$, where the index $a$ spans the rows of the corresponding
Young diagram, and let $r(R)$ denote the $U(1)$ charge of $R$. Then
the second Casimir invariant of $R$ can be conveniently written as
\beq C_2(R)=\sum_{a=1}^N\, \left(\hat
n_a(R)+r(R)-a-\mbox{$\frac{N-1}{2}$}\right)^2=\left\{\begin{array}{ll}
\displaystyle\sum_{a=1}^N\,
n_a(R)^2 &\ \ \ \ \
\mathrm{for\ }N\mathrm{\ odd} \ , \\  \\
\displaystyle\sum_{a=1}^N\,\left(n_a(R)-\mbox{$
\frac{1}{2}$}\right)^2 &\ \ \ \ \
\mathrm{for\ }N\mathrm{\ even \ , }
\end{array}\right.
\eeq
where in the second equality we have absorbed an irrelevant shift into
the weight integers $\mbf n(R)\in\mathds{Z}^N$. (We have also dropped
an overall factor depending only on $N$.) Note that the trivial
representation $R=0$ has weight $\mbf n(0)=\mbf0$. Throughout we will
assume that the rank $N$ is
odd. This restriction is not necessary but it will simplify some of
our analysis in the following. The correlators $S_{R_i, R_{i+1}}$
appearing in (\ref{Zpq}) arise from the gluing of disks and
annuli to build the necklace of $\ell$ spheres, and they are given by
\beq
S_{R,Q}=\sum_{w\in S_N}\,\varepsilon(w)~\e^{-\frac{g_{\rm YM}^2}
{2}\,w(\mbf n(R)+\mbf\rho)\cdot(\mbf n(Q)+\mbf\rho)} \ .
\label{Smatrixdef}\eeq This operator is related to the modular
S-matrix of the $U(N)$ WZW model in the Verlinde basis (see
Section~\ref{CSLpqPartFn}). Here $\mbf\rho$ is the Weyl vector of
$U(N)$ (the half sum of positive roots) whose components are given by
\beq
\rho_a=\mbox{$\frac{N-2a+1}{2}$} \ , \eeq and the
elements $w$ of the Weyl group $S_N$ of $U(N)$ act by permuting the
entries of $N$-vectors with sign $\varepsilon(w)$.

\subsection{Semi-Classical Expansion\label{InstExp}}

The Poisson resummation of (\ref{Zpq}) has a natural interpretation as
an expansion of the q-deformed gauge theory into a sum over
classical solutions~\cite{Arsiwalla:2005jb}--\cite{Caporaso:2005ta}.
After some trivial manipulations and dropping of an overall irrelevant
normalization, we can recast the partition function in the form
\bea
{Z}_{U(N)}^{q{\rm YM}}\big(X(p,q)\,,\,g_{\rm YM}^2\big)
&=&\sum_{w\in S_N}\,\varepsilon(w)~\sum_{\mbf
  n_1,\dots,\mbf n_\ell\in \mathds{Z}^N}\,
\e^{-\frac{g_{\rm YM}^2}{4}\,\sum_i\, e_i\,{\mbf n}_i^2-
\frac{g_{\rm
    YM}^2}{2}\,\sum_{i<j}\,\mbf{n}_i\cdot\mbf{n}_j+\ii\sum_i\,
\mbf{f}_i\cdot\mbf n_i}\nonumber\\ 
&& \qquad\qquad\qquad\qquad\qquad\qquad \times~
\e^{-\frac{g_{\rm YM}^2}2\,(\mbf{\rho}\cdot\mbf{n}_1-w(
\mbf{\rho})\cdot\mbf{n}_\R)} \ ,
\label{Zpq1}\eea
where $\mbf{f}_i=\theta_i\,(1,1,\dots,1)$. Each integer vector $\mbf
n_i$, $i=1,\dots,\ell$ classifies one of the original irreducible
$U(N)$ representations $R_i$ appearing in (\ref{Zpq}). The quadratic form
in the exponent of (\ref{Zpq1}) can be succinctly rewritten
in terms of the Cartan matrix (\ref{InterMatrix}) to get
\beq
{Z}_{U(N)}^{q{\rm YM}}\big(X(p,q)\,,\,g_{\rm YM}^2\big)
=\sum_{w\in S_N}\,\varepsilon(w)~
\sum_{\mbf n_1,\dots,\mbf n_\ell\in \mathds{Z}^N}\,
\e^{-\frac{g_{\rm YM}^2}{4}\,C_{ij} \,\mbf n_i\cdot\mbf{n}_j+
  \ii\mbf f_i\cdot \mbf n_i}~\e^{
-\frac{g_{\rm YM}^2}2\,(\mbf{\rho}\cdot\mbf{n}_1-w(\mbf{\rho})\cdot
\mbf{n}_{\R})}
\eeq
with an implicit sum over repeated indices.

The desired modular inversion is now realized through an elementary
gaussian integration and one finds
\beq
\begin{split}
{Z}_{U(N)}^{q{\rm YM}}\big(X(p,q)\,,\,g_{\rm YM}^2\big)
=&\left({\frac{4\pi}{g_{\rm YM}^2}}\right)^{N \,\R/2}\,
\frac{1}{\det(C)^{N/2}}\\ & \times\,\sum_{\mbf{m}_1,\dots,\mbf{m}_{\R}\in
\mathds{Z}^{N}}\,\e^{-\frac{4\pi^2}{g_{\rm YM}^2}\, (C^{-1})^{ij}\,
(\mbf{m}_i+\frac{\mbf f_i}{2\pi})\cdot
(\mbf{m}_j+\frac{\mbf f_j}{2\pi})}\\
&\times\,\sum_{w\in S_N}\,\varepsilon(w)~
\e^{2\pi \ii (\mbf{m}_i +\frac{\mbf f_i}{2\pi})\cdot
((C^{-1})^{1i}\,\mbf{\rho} +(C^{-1})^{i\R}\,w(\mbf{\rho}))}\\
&\qquad\qquad\times~\e^{\frac{g_{\rm YM}^2}{4} \,
((C^{-1})^{1 1}\,\mbf{\rho}^2
 +2 (C^{-1})^{1 \R}\,w(\mbf{\rho})\cdot \mbf{\rho}+
(C^{-1})^{\R\R}\,w(\mbf{\rho})^2)} \ .
\end{split}
\eeq
This expression can be simplified by exploiting the explicit form for
the inverse of the Cartan matrix provided in Appendix~\ref{Bappendix},
where the definitions of the integers $q_i$ and $p_i$ appearing below
may be found. We obtain
\beq
\begin{split}
{Z}_{U(N)}^{q{\rm YM}}\big(X(p,q)\,,\,g_{\rm YM}^2\big)=&~\mathcal{N}\,
\sum_{\mbf{m}_1,\dots,\mbf{m}_\ell\in \mathds{Z}^{N}}\,
\e^{-\frac{4\pi^2}{g_{\rm YM}^2}\, (C^{-1})^{ij}\,
\mbf{m}_i\cdot\mbf m_j-\frac{4\pi}{g_{\rm YM}^2}\,\mbf m_i\cdot \mbf f_i
}\\ & \times\, \sum_{w\in S_N}\,\varepsilon(w)~
\e^{\frac{2\pi \ii}{p} \, q_i\,\mbf m_i\cdot
(q\,\mbf\rho+w(\mbf\rho))+\frac{g_{\rm YM}^2}{2p}\,w(
\mbf{\rho})\cdot \mbf{\rho}}
\end{split}
\label{Zpqinstsimpl}\eeq
where
\beq
\mathcal{N}:=\left({\frac{4\pi}{g_{\rm YM}^2}}\right)^{N\, \R/2}\,
p^{-N/2}~\e^{\frac{\ii N}{p}\,(p_i+q_i)\,\theta_i +
\frac{g_{\rm YM}^2}{24 p}\, (N^3-N)\,(q+q^\prime\,)-
\frac{N}{g_{\rm YM}^2}\, (C^{-1})^{ij} \,\theta_i\,\theta_j} \ .
\eeq

The sum over permutations $w$ in (\ref{Zpqinstsimpl}) does not depend
on the instanton numbers $\mbf m_i$ individually, but rather only
on the linear combination $q_i\,\mbf m_i$.
This special dependence suggests the change of variables
\beq
\mbf s_1=q_i\,\mbf m_i \qquad \mbox{and} \qquad \mbf s_j=\mbf m_j \ \
\mathrm{for}\ \ j=2,\dots,\R
\eeq
with $q_1=1$ (see Appendix~\ref{Bappendix}) and
\beq
\mbf m_1=\mbf s_1- \sum_{i=2}^\R\,  q_i\, \mbf s_i \ .
\eeq
Then the partition function (\ref{Zpqinstsimpl}) takes the form
\bea
{Z}_{U(N)}^{q{\rm YM}}\big(X(p,q)\,,\,g_{\rm YM}^2\big)&=&\mathcal{N}\,
\sum_{{\mbf{s}}_1,\dots,{\mbf{s}}_\ell\in \mathds{Z}^{N}}\,
\e^{-\frac{4\pi^2}{g_{\rm YM}^2}\,
  q_j\, h_{i}\,\mbf{s}_i\cdot\mbf{s}_j-\frac{4}{g_{\rm YM}^2\, p}\,
\sum_{j=2}^k\, q_j\,\mbf{s}_1\cdot \mbf f_j}~
\e^{-\frac{4\pi^2}{g_{\rm YM}^2}\frac{q\,\mbf{s}_1^2}{p} }\nonumber\\
&& \times\,\sum_{w\in S_N}\,\varepsilon(w)~
\e^{\frac{2\pi \ii}{p}\, \mbf{s}_1\cdot(q\,\mbf \rho+w(\mbf\rho))+
\frac{g_{\rm YM}^2}{2p}\,w(\mbf{\rho})\cdot \mbf{\rho}} \ ,
\label{Zpqinstchange}\eea
where the integers $h_i$, $i=1,\dots,\ell$ are defined in
Appendix~\ref{Bappendix}. The sum over permutations in
(\ref{Zpqinstchange}) now depends only on the single integer
$\mbf{s}_1$. Since the Weyl vector $\mbf\rho$ is integer-valued for
$N$ odd, the dependence on $\mbf{s}_1$ is periodic with period $p$. It
is natural then to decompose the sum over $\mbf{s}_1$ in two separate
steps. First we sum over $\mbf{s}_1$ modulo $p$, and then we sum over
all integer multiples of $p$. This is achieved through the change of
variable
\beq
\mbf{s}_1~\longrightarrow~\mbf m+p\,\mbf{s}_1
\eeq
with $\mbf m\in\zed_p^N$ and $\mbf{s}_1\in\zed^N$.

After this final change of variable, we can recast the partition
function (\ref{Zpqinstchange}) in the form
\bea
{Z}_{U(N)}^{q{\rm YM}}\big(X(p,q)\,,\,g_{\rm YM}^2\big)&=&\mathcal{N}\,
\sum_{{\mbf{s}}_1,\dots,{\mbf{s}}_\R\in \mathds{Z}^{N}}~
\sum_{\mbf m\in \mathds{Z}^N_p}\,\e^{-\frac{4\pi^2}{g_{\rm YM}^2}\,
 G_{ij}\,\mbf{s}_i\cdot\mbf{s}_j-\frac{4}{g_{\rm YM}^2\, p}\,
\sum_{j=2}^k\, q_j\,\mbf f_j\cdot(p\, \mbf{s}_1+ \mbf m)}\nonumber\\
&& \qquad\qquad\qquad\qquad \times~
\e^{ -\frac{8\pi^2}{g_{\rm YM}^2}\,{q\,\mbf{s}_1\cdot\mbf m}}~
Z_{U(N)}^{\rm CS}\big(L(p,q)\,,\,\mbf m\big) \ ,
\label{Finaltheta}\eea
where the symmetric $\ell\times\ell$ integer-valued matrix $G_{ij}$ is
defined in Appendix~\ref{Bappendix} and
\beq
Z_{U(N)}^{\rm CS}\big(L(p,q)\,,\,\mbf m\big)=
\e^{-\frac{4\pi^2}{g_{\rm YM}^2}\,\frac{q\,\mbf m^2}{p} }~
\sum_{w\in S_N}\,\varepsilon(w)~
\e^{\frac{2\pi \ii}{p}\,\mbf  m\cdot(q\,\mbf\rho+w(\mbf\rho))+
\frac{g_{\rm YM}^2}{2p}\,w(\mbf{\rho})\cdot \mbf{\rho}} \ .
\label{ZCSUNLpqm}\eeq
If we now set
\beq
g_{\rm YM}^2=\frac{4\pi \ii}{k+N}
\label{gYMkCS}\eeq
then (\ref{ZCSUNLpqm}) can be identified as the partition function of
$U(N)$ Chern-Simons gauge theory with level $k\in\nat_0$ on the Lens
space $L(p,q)$ in the background of the flat connection defined by the
torsion vector $\mbf m\in\zed_p^N$. We will derive this explicitly in
Section~\ref{CSLpq}. The relationship between q-deformed Yang-Mills
theory and Chern-Simons theory is not surprising and was anticipated
by~\cite{Aganagic:2004js,Blau:2006gh,Beasley:2005vf}. In this paper we
extend this correspondence very explicitly to the generic chain of
exceptional divisors of the Hirzebruch-Jung spaces $X(p,q)$ whose
boundaries are the more general Lens spaces $L(p,q)$.

\subsection{Emergence of Four-Dimensional Instantons\label{2Dto4D}}

In~\cite{Vafa:2004qa,Aganagic:2004js,Aganagic2:2005} the expression
(\ref{Zpq}) is conjectured to be the partition function of the
topologically twisted $\mathcal{N}=4$ supersymmetric Yang-Mills theory
on the four-dimensional toric manifold $X(p,q)$ obtained by blowing up
the $A_{p,q}$ singularity. However, as is manifest from the form
(\ref{Finaltheta}), this interpretation is {\it a priori}
difficult. Although there is a sum over $\ell$ vectors
$\mbf{s}_i\in\mathds{Z}^N$ playing the putative role of instanton
numbers, this contribution is accompanied in (\ref{ZCSUNLpqm}) by a
second sum over permutations $w$ of $N$ elements which is {\it
  perturbative} in its dependence on the Yang-Mills coupling
constant. Such terms are interpreted as fluctuations around the given instanton
background. However, in the topologically twisted ${\cal N}=4$ gauge
theory, contributions of this sort are absent because the partition
function simply computes the Euler characteristic of the instanton moduli
space. Similar problems with interpreting two-dimensional gauge theory
partition functions as generating functions for instanton counting
were noticed in~\cite{Aganagic2:2005,Caporaso:2006kk}.

It is clear from eq.~(\ref{Finaltheta}) that the two-dimensional gauge
theory treats the asymptotic values of the gauge fields on the
boundary $L(p,q)$ of $X(p,q)$ as dynamical quantities, whose evolution
is governed by the Chern-Simons action on the Lens space
$L(p,q)$. This is in marked contrast with what happens in a typical
instanton computation, whereby the gauge fields on the boundary are
fixed quantities and the partition function is simply obtained by
summing over the admissible boundary conditions. In other words, in
order to make contact with the four-dimensional instanton
computations, it appears natural to drop the perturbative contribution
to the Chern-Simons partition function (\ref{ZCSUNLpqm}) from
(\ref{Finaltheta}) and keep only its classical part. This modifies
(\ref{Finaltheta}) to the partition function
\bea
{\mathcal{Z}}_{X(p,q)}^{U(N)}&=&\mathcal{N}\,
\sum_{{\mbf{s}}_1,\dots,{\mbf{s}}_\R\in \mathds{Z}^{N}}~
\sum_{\mbf m\in{\mathds{Z}^N_p}}\,\e^{-\frac{4\pi^2}{g_{\rm YM}^2}\,
 G_{ij}\,\mbf{s}_i\cdot\mbf{s}_j-\frac{4}{g_{\rm YM}^2\, p}\,
\sum_{j=2}^\R\, q_j\,\mbf f_j\cdot(p \,\mbf{s}_1+ \mbf m)}
\nonumber\\ && \qquad\qquad\qquad\qquad\qquad \times~
\e^{-\frac{4\pi^2}{g_{\rm YM}^2}\,\frac{q\,\mbf m^2}{p} -
\frac{8\pi^2}{g_{\rm YM}^2}\,{q\,\mbf{s}_1\cdot\mbf m}} \ .
\label{Final0}\eea
After evaluating the lattice Gauss sum over $\mbf m\in\zed_p^N$, we
obtain finally the result
\beq
\label{Final1}
{\mathcal{Z}}_{X(p,q)}^{U(N)}=\mathcal{N}\,
\sum_{\mbf{u}_1,\dots,\mbf{u}_\R\in \mathds{Z}^{N}}\,
\e^{-\frac{4\pi^2}{g_{\rm YM}^2}\,
(C^{-1})^{ij}\,\mbf u_i\cdot\mbf u_j -\mbf z^i\cdot\mbf u_i}
\eeq
where we have identified $\mbf z^i= 4\,(C^{-1})^{ij}\,\mbf f_j/g^2_{\rm YM}$
in terms of the chemical potentials for D2-branes wrapped on the
exceptional divisors of $X(p,q)$.

The expression (\ref{Final1}) has a nice interpretation in terms of
instanton counting in the ${\cal N}=4$ topologically twisted
Yang-Mills theory. Apart from the trivial overall normalization
factor, it corresponds precisely to the contribution of fractional
instantons to the ${\cal N}=4$ partition function, as provided
explicitly in~\cite{fcr} for the family $X(p,q)$ of toric
four-manifolds. This identification can also be made by recognizing
that the exponent in (\ref{Final1}) is exactly the classical action
for fractional instantons, as we showed in Section~\ref{InstSect} (see
eq.~(\ref{frac0})). The
fact that the two-dimensional gauge theory naturally captures the
contributions of fractional instantons can be understood by noting
that they are the pullbacks to $X(p,q)$ of the two-dimensional
classical solutions. There are in fact bijective correspondences
between two-dimensional instantons on a certain orbifold of $\PP^1$,
flat three-dimensional connections on the boundary $L(p,q)$, and
fractional instantons on the four-manifold $X(p,q)$ as illustrated by
the discussion of Section~\ref{InstSect} and shown in detail in
Section~\ref{CSLpq} below. Moreover, the expression (\ref{Final0})
yields an even more refined formula providing the contributions of
each topological sector of fractional instantons with fixed holonomy
$\mbf m\in\zed_p^N$ at infinity from the finite action requirement
that the gauge fields be asymptotically flat.

The contributions from regular instantons are more elusive, since they
can move freely on the whole non-compact four-dimensional space, and
require in general some sort of regularization procedure. In the
q-deformed gauge theory, these ambiguities are most evident on the
toric manifold $\mathds{C}^2$, where the fractional instanton
contribution is absent. In this case, the gluing rules of
Section~\ref{Sewing} above yield a divergent amplitude which must be
suitably regularized. To elucidate further the structure of our result and
compare with existing results in the literature, in the remainder of
this section we will look more closely at the two extreme cases
${{\cal O}_{\PP^1}(-p)}$ ($q=1$) and ALE spaces ($q=p-1$).

\subsection{Example: Line Bundles over $\PP^1$\label{P1Example}}

The limiting case $X(p,1)$ is the total space of the holomorphic line
bundle $\mathcal{O}_{\PP^1}(-p)$ of degree $p$ over $\PP^1$. In this
case the Cartan matrix (\ref{InterMatrix}) has just one element
$e_1=p$. The partition function (\ref{Zpq}) is that of q-deformed
Yang-Mills theory on the sphere, whose instanton expansion was
worked out explicitly in~\cite{Caporaso:2005ta} and written for
$\theta_1=0$ in the compact form
\beq
{Z}_{U(N)}^{q{\rm YM}}\big({\mathcal{O}_{\PP^1}(-p)}\,,\,
g_{\rm YM}^2\big)= \sum_{\stackrel{\scriptstyle\mbf N\in\nat_0^p}
{\scriptstyle\sum_k\,N_k=N}}~\prod_{k=0}^{p-1}\,\frac{\theta_3\left(
\left.\frac{4\pi\ii p}{g_{\rm YM}^2} \right|
 \frac{4\pi\ii k}{g_{\rm YM}^2}\right)^{N_k}}{N_k!}~Z_{U(N)}^{\rm
CS}\bigl(L(p,1)\,,\,\mbf N\bigr) \ ,
\label{ZP1ZCS}\eeq
where
\beq
\theta_3(\tau|z)=\sum_{m\in \mathbb{Z}}\e^{\pi \ii\tau
\,{{m}}^2 + 2 \pi \ii {m}\,{z}}
\label{Jacobi3def}\eeq
is a Jacobi-Erderlyi theta-function. Here
\beq
Z_{U(N)}^{\rm CS}\bigl(L(p,1)\,,\,\mbf N\bigr)=\exp\left(-\frac{4\pi^2}
{g_{\rm YM}^2\,p}\,\sum_{m=0}^{p-1}\, N_m \,m^2\right)~
{\cal W}_p^{\mathrm{inst}}(\,\underbrace{0,\dots,0}_{N_0},\dots,
 \underbrace{p-1,\dots,p-1}_{N_{p-1}}\,) \ , \label{LS}
\eeq
with
\beq
{\cal W}_p^{\mathrm{inst}}(\mbf s)=\left(\frac{4\pi}{g_{\rm YM}^2\,p}
\right)^{N/2}~\e^{-\frac{g_{\rm YM}^2\,(N^3-N)}{12p}}~
\sum_{w\in S_N}\,\varepsilon(w)~
\e^{\frac{2\pi \ii}{p}\,\mbf s\cdot(\mbf\rho+w(\mbf\rho))+
\frac{g_{\rm YM}^2}{4p}  ({\mbf\rho}^2
 +2w({\mbf\rho})\cdot {\mbf\rho})} \ , \label{LSfluct}
\eeq
is simply the partition function of $U(N)$ Chern-Simons
gauge theory on the Lens space $L(p,1)$, the boundary of the total
space of the line bundle $\mathcal{O}_{\PP^1}(-p)$, for the vacuum
contribution corresponding to the $p$-component partition $\mbf
N\in\nat_0^p$ of $N$, as computed
in~\cite{Marino:2002fk,Aganagic:2002wv} and in Section~\ref{CSLpq}
below.

In this case, the only surviving sum in (\ref{ZP1ZCS}) is carried over
ordered partitions $\mbf N$ of $N$ into $p$ parts. This ensures that
we are summing over gauge inequivalent flat connections on the
boundary of $\mathcal{O}_{\PP^1}(-p)$. If we now drop the perturbative
contribution in (\ref{LSfluct}), we immediately find
\bea
{\cal Z}^{U(N)}_{\mathcal{O}_{\PP^1}(-p)}&=&
\mathcal{N}\,\sum_{\stackrel{\scriptstyle\mbf N\in\nat_0^p}
{\scriptstyle\sum_k\,N_k=N}}\,
\exp\left(-\frac{4\pi^2}{g_{\rm YM}^2\,p}\,
\sum_{m=0}^{p-1}\, N_m\, m^2\right)~
\prod_{k=0}^{p-1}\,\frac{\theta_3\left(
\left.\frac{4\pi\ii p}{g_{\rm YM}^2} \right|
 \frac{4\pi\ii k}{g_{\rm YM}^2}\right)^{N_k}}{N_k!}\nonumber\\[4pt]
 &=&\frac{1}{N!}\,{\cal N}\,\left[~\sum_{k=0}^{p-1}\,\e^{-\frac{4\pi^2\,k^2
}{g^2_{\rm YM}\,p}}~\theta_3\left(\mbox{$
\left.\frac{4\pi\ii p}{g^2_{\rm YM}} \right|
 \frac{4\pi\ii k}{g^2_{\rm YM}}$}\right)\right]^N\nonumber\\[4pt]
&=& \frac{1}{N!}\,{\cal N}\,\left[~
\sum_{k=0}^{p-1}~\sum_{m\in \mathbb{Z}}\,\e^{-\frac{4\pi^2}
{g^2_{\rm YM}\,p}\,(k+p\,m)^2}\right]^N=
\frac{1}{N!}\,{\cal N}~\theta_3\left (\mbox{$\left.
\frac{4\pi\ii}{g_{\rm YM}^2\,p}\right|0$}\right)^N \ .
\label{LS1}\eea
This coincides with the contribution, derived in~\cite{fcr}, of
fractional instantons to the partition function of $\mathcal{N}=4$
gauge theory on $\mathcal{O}_{\PP^1}(-p)$ in the absence of
D2-branes.

\subsection{Example: ALE Spaces\label{ALEExample}}

The partition function of q-deformed Yang-Mills theory on the ALE
spaces $A_k$ was first computed in~\cite{Aganagic2:2005} by embedding
this space into the local Calabi-Yau threefold
$A_k\times\mathds{C}$. This threefold can be thought of as the limit
of the usual ALE fibration over $\mathds{P}^1$ as the area of the base
$\mathds{P}^1$ becomes infinite. Setting $p=k+1$ and $q=k$, the
instanton representation of the partition function is given by
\bea
{Z}_{U(N)}^{q{\rm YM}}\big(A_k\,,\,g_{\rm YM}^2\big)
&=&\left({\frac{4\pi}{g_{\rm YM}^2}}\right)^{N\,k/2}\,
\frac{1}{(k+1)^{N/2}} \\ &&\times\,
\sum_{\mbf s_0,{\mbf s}_1,\dots,{\mbf s}_{k-1}\in \mathds{Z}^{N}}~
\sum_{\mbf m\in \zed^N_p}\,\e^{-\frac{4\pi^2}{g_{\rm YM}^2}\,
\mathcal{ A}^{ij}\,\mbf s_i\cdot \mbf s_j-
\frac{8\pi^2}{g_{\rm YM}^2}\, \sum_{j} \,(k-j) \,\mbf s_j\cdot
\mbf m}\nonumber \\
&&\times\, \e^{-\frac{4\pi^2}{g_{\rm YM}^2}\,\frac{k}{k+1}\,
\mbf m^2}\,\sum_{w\in S_N}\,\varepsilon(w)~
\e^{\frac{2\pi \ii k }{k+1}\,\mbf m\cdot(w(\mbf\rho)-
\mbf\rho)+\frac{g_{\rm YM}^2}{2(k+1)}\,  (k\,{\mbf\rho}^2
 +w({\mbf\rho})\cdot {\mbf\rho})}\nonumber
\eea
where the symmetric $k\times k$ matrix elements $\mathcal{A}^{ij}$ for
$i,j=1,\dots,k-1$ are given by
\beq
{\mathcal{A}}^{ij}=\big(
C^{-1}\big)^{ij}+\frac{1}{2(k+1)}\,\Big[\big(k+1-i\big)\,\big(k\,(k+1-j)-2
j\big)+\big(k+1-j\big)\,\big(k\,(k+1-i)-2 i\big)\Big] \ ,
\eeq
while ${\mathcal{A}}^{0j}=(k+1)\,(k-j) $ for any $j=0,1,\dots,k-1$. To
avoid an overly cumbersome expression we have set all $\theta$-angles
$\theta_i=0$. We have also used the fact that the Cartan matrix
for ALE spaces coincides with the Cartan matrix for the $A_k$ Dynkin
diagram
\beq
C=\begin{pmatrix} 2 & -1 & 0 & \cdots & 0 \\
-1 &2  & -1& \cdots & 0 \\
0 &-1 & 2&\cdots & 0 \\
\vdots &\vdots & \vdots &\ddots&\vdots\\
0&0&0&\cdots&2
\end{pmatrix} \ .
\eeq
Dropping the perturbative contribution again we arrive at
\beq
\begin{split}
{\cal Z}_{A_k}^{U(N)}
&={\cal N}\,\sum_{\mbf{u}_1,\dots,{\mbf u}_k\in
    \mathds{Z}^{N}}\,\e^{-\frac{4\pi^2}{g_{\rm YM}^2}\,
(C^{-1})^{ij}\,\mbf u_i\cdot\mbf u_j} \ ,
\end{split}
\eeq
which is exactly the contribution of fractional instantons to the
${\cal N}=4$ Yang-Mills partition function, derived in~\cite{fcr} via
explicit instanton computations, given by the $N$-th power of
(\ref{frac0}) for $z^i=0$.

\section{Chern-Simons Gauge Theory on Lens Spaces\label{CSLpq}}

In this section we will describe in some detail the nonabelian
localization of $U(N)$ Chern-Simons gauge theory on the generic
three-dimensional Lens spaces $L(p,q)$, and thereby prove some of the
assertions made in the previous section. The quantum gauge theory is
defined by the path integral
\begin{equation}
\mathcal{Z}_{U(N)}^{\rm CS}\big(L(p,q) \,,\, k\big) = \int\, \DD A ~
\e^{-S_{U(N)}^{\rm CS}(A)}
\label{CSpartfn}\end{equation}
where
\beq
S_{U(N)}^{\rm CS}(A)=
\frac{\ii k}{4 \pi }\, \int_{L(p,q)}\, \Tr\left(A \wedge
    \mathrm{d} A + \mbox{$\frac{2}{3}$}\, A \wedge A \wedge A \right)
\label{CSaction}\eeq
is the Chern-Simons action evaluated on a connection $A$ of a
principal $U(N)$ bundle over $L(p,q)$. As is
well-known~\cite{Beasley:2005vf}, the partition function
(\ref{CSpartfn}) is given exactly by its semi-classical
approximation. For this, one takes into account the one-loop radiative
correction $k\to k+N$ and sums over all critical points of the
Chern-Simons action (\ref{CSaction}), which are simply the flat $U(N)$
gauge connections on $L(p,q)$. The purpose of carrying out this
calculation explicitly is two-fold. Firstly, we will correctly
identify for the first time the individual flat connection
contributions to the semi-classical expansion of (\ref{CSpartfn}),
generalizing the results of~\cite{Marino:2002fk,Aganagic:2002wv} to
the generic Lens spaces $L(p,q)$ for all $1\leq q<p$ and justifying
our identification (\ref{ZCSUNLpqm}). Secondly, in the course of this
calculation we will construct the explicit mapping between
two-dimensional Yang-Mills instantons on a $\PP^1$ orbifold, flat
connections on $L(p,q)$, and fractional instantons on $X(p,q)$, which
is the crux of the final results of the previous section.

\subsection{Classical Solutions\label{FlatLpq}}

We begin by constructing the flat connections on $L(p,q)$
explicitly. For this, it is convenient to realize the Lens space
$L(p,q)$ as a Seifert fibration over the two-sphere~\cite{Furuta} (see
also~\cite{Beasley:2005vf}). The base is
described by a projective line $\PP^1$ with an arbitrarily chosen
marked point at which the coordinate neighbourhood is modelled on
$\complex/\zed_p$, with the cyclic group acting on the local chart
coordinate $z$ as $z\mapsto\e^{2\pi\ii/p}\,z$. We construct a line
V-bundle ${\cal L}(p,q)$ over this $\PP^1$ orbifold such that the
local trivialization over the orbifold point is modelled by
$\complex^2/\zed_p$, where $\zed_p$ acts on the local
coordinates $(z,w)$ of the base and fibre exactly as
in~(\ref{gamma}). The Lens space $L(p,q)$ may then be described as
the total space of the associated unit circle bundle $\sphere({\cal
  L}(p,q))$. Since $p$ and $q$ are relatively prime, this
construction realizes $L(p,q)$ as the quotient of the three-sphere
$\sphere^3$ by the free $\zed_p$-action (\ref{gamma}), where
$\sphere^3$ is regarded as the unit sphere in $\complex^2$. Since
$\pi_1(\sphere^3)=0$, it follows that the fundamental group of the
Lens space is simply
\beq
\pi_1\big(L(p,q)\big)=\pi_0\big(\zed_p\big)=\zed_p
\label{pi1Lens}\eeq
and it is generated by the noncontractible loop encircling the
orbifold point on the base $\PP^1$. Moreover, the Chern class of the
line V-bundle over $\PP^1$ describing $L(p,q)$ is
\beq
c_1\big({\cal L}(p,q)\big)=\frac qp \ .
\label{c1Lpq}\eeq
This class cancels the local delta-function curvature at the marked
point of $\PP^1$ to ensure that the total degree of the fibration
is~$0$.

Gauge equivalence classes of flat $U(N)$ connections on $L(p,q)$ are
in one-to-one correspondence with conjugacy classes of homomorphisms
$\rho$ from the fundamental group (\ref{pi1Lens}) to $U(N)$, i.e. with
$N$-dimensional unitary representations of $\zed_p$. The image of
$\rho$ in $U(N)$ decomposes into $N_m$ copies of the $m$-th
one-dimensional irreducible representation of $\zed_p$, where
$m=0,1,\dots,p-1$, and any representation $\rho$ lives in the maximal
torus $U(1)^N\subset U(N)$ with
\beq
N=\sum_{m=0}^{p-1}\,N_m \ .
\label{partpconstr}\eeq
It follows that there is a one-to-one correspondence between flat
$U(N)$ gauge connections on $L(p,q)$ and $p$-component partitions
$\mbf N\in\nat_0^p$ of the rank $N$. Moreover, any such
connection defines a central element of the Lie algebra of $U(N)$.

The isomorphism class $[{\cal T}]$ of the tautological line bundle
over $\PP^1$ is the generator of $H^1(\PP^1,U(1))\cong
H^2(\PP^1,\zed)\cong\zed$. Since $H^1(\PP^1,\zed)=0$, it follows from
the Thom-Gysin exact sequence for circle bundles~\cite{Furuta} that
$H^2(L(p,q),\zed)=H^2(\PP^1,\zed)/\langle[{\cal
  T}^{p}]\rangle\cong\zed_p$. This means that all unitary vector
bundles over $L(p,q)$ have $p$-torsion magnetic charges (Chern
classes) $m$, and that all such torsion bundles over $L(p,q)$ are
pullbacks of ordinary bundles over $\PP^1$ under the bundle projection
$\pi:\sphere({\cal L}(p,q))\to\PP^1$. As we now explicitly
demonstrate, this implies that every flat connection on $L(p,q)$ is
the pullback of a configuration of Dirac monopoles on the
sphere $\PP^1$. Extending the pullback to the bulk $X(p,q)$ is then in
agreement with the construction of fractional instantons given in
Section~\ref{InstSect}.

The critical points of the Yang-Mills action functional
$\frac1{2g^2}\,\int_{\PP^1}\,\Tr(F_a\wedge{}^*F_a)$ are the $U(N)$
gauge connections $a$ satisfying $\dd_a{}^*F_a=0$. These are the
connections with constant central curvature. On the two-sphere every
constant curvature bundle is (up to isomorphism) a sum of line
bundles. There is thus a one-to-one correspondence between Yang-Mills
connections of a principal $U(N)$-bundle $\cal P$ of degree $m$ over $\PP^1$
and non-increasing sequences of integers $\mbf m\in\zed^r$, of
respective multiplicities $\mbf N\in\nat_0^r$, with
\beq
m=\sum_{i=1}^r\,m_i \qquad \mbox{and} \qquad N=\sum_{i=1}^r\,N_i \ .
\label{YMpartitions}\eeq
On the sphere $\PP^1$, each such connection is gauge equivalent to the
connection $a_0(\mbf m,\mbf N)=\bigoplus_i\,a^{(m_i)}~\id_{N_i}$, where
$a^{(m_i)}$ is the monopole potential of magnetic charge $m_i$ and the
$i$-th block is an abelian connection on the bundle $({\cal
  T}^{m_i})^{\oplus N_i}$. The curvature of this connection is given
by
\beq
F_{a_0}(\mbf m,\mbf N)=\dd a_0(\mbf m,\mbf N)=
\bigoplus_{i=1}^r\,2\pi\,m_i~\id_{N_i}
\otimes\omega_{\PP^1} \ ,
\label{centralcurvYM}\eeq
where $\omega_{\PP^1}$ is the symplectic two-form on $\PP^1$
normalized to unit volume
\beq
\int_{\PP^1}\,\omega_{\PP^1}=1 \ .
\label{omegaPP1int}\eeq

The monopole connection of course has trivial monodromies around
arbitrary smooth points on $\PP^1$. To take into account the
orbifolding of $\PP^1$ required to define the Lens space as a Seifert
manifold, we need a connection which has non-trivial monodromy
$\e^{2\pi\ii q/p}$ about the given fixed marked point on
$\PP^1$. Since the choice of orbifold point is arbitrary, we may thus
define
\beq
a(\mbf m,\mbf N):=\mbox{$\frac qp$}\,a_0(\mbf m,\mbf N)
\qquad \mbox{and} \qquad F_a(\mbf m,\mbf N)=\dd a(\mbf m,\mbf N)=
\mbox{$\frac qp$}\,F_{a_0}(\mbf m,\mbf N)
\label{centralmonod}\eeq
with the requisite monodromy. In particular, the Chern class
(\ref{c1Lpq}) has a Chern-Weil description in terms of smooth
curvature in the bulk of the $\PP^1$ orbifold as $c_1({\cal
  L}(p,q))=\frac1{2\pi}\,\int_{\PP^1}\,F_a(1,1)$. The holonomy of
this abelian gauge connection depends only on the values of the monopole
numbers $m_i$ mod~$p$. By a trivial rearrangement, we will denote by
$0\leq N_m\leq N$ the multiplicity of the degree~$m$ monopole bundle
${\cal T}^{m}$ for the torsion magnetic charges $m=0,1,\dots,p-1$
alluded to earlier and hence drop the labels $\mbf m$ from the
notation above.

To describe the pullback of these gauge fields to $L(p,q)$, we use
(\ref{c1Lpq}) to introduce a connection $\kappa$ on the principal
$U(1)$-bundle $\pi:\sphere({\cal L}(p,q))\to\PP^1$ whose curvature is
given by
\beq
\dd\kappa=\mbox{$\frac qp$}\,\pi^*(\omega_{\PP^1}) \ .
\label{kappacurv}\eeq
The integral of $\kappa$ over a generic fibre of the Seifert fibration
is given by~\cite{Blau:2006gh,Beasley:2005vf}
\beq
\oint_{\sphere^1}\,\kappa=1 \ ,
\label{intkappa}\eeq
while from (\ref{omegaPP1int}), (\ref{kappacurv}) and (\ref{intkappa})
it follows that the Chern class (\ref{c1Lpq}) of the line V-bundle ${\cal
  L}(p,q)$ can be computed from the integral
\beq
\int_{L(p,q)}\,\kappa\wedge\dd\kappa=\frac qp \ .
\label{c1integral}\eeq
We can now compute the pullback of the curvature in
(\ref{centralmonod}) as
\beq
F_A\big(\mbf N\big):=
\pi^*\big(F_a(\mbf N)\big)=\bigoplus_{m=0}^{p-1}\,\frac{2\pi\,m\,q}p~
\id_{N_m}\otimes\pi^*\big(\omega_{\PP^1}\big)=\bigoplus_{m=0}^{p-1}\,
2\pi\,m~\id_{N_m}\otimes\dd\kappa \ ,
\label{Fapullback}\eeq
from which we may identify the pullback of the Yang-Mills instanton on
the $\PP^1$ orbifold up to gauge transformation as
\beq
A(\mbf N)=\bigoplus_{m=0}^{p-1}\,
2\pi\,m~\id_{N_m}\otimes\kappa \ .
\label{YMinstpullback}\eeq
Note that from (\ref{intkappa}) it follows that the connection
(\ref{YMinstpullback}) has trivial holonomy along any $\sphere^1$
fibre of the Seifert manifold, $\exp(\ii\oint_{\sphere^1}\,A(\mbf
N))=\id_N$, as required since all fibre loops are contractible in
$L(p,q)$ and the only non-trivial elements of (\ref{pi1Lens}) arise
from loops which wind around the marked point of the base
$\PP^1$. Moreover, if ${\cal P}\to\PP^1$ is the irreducible $U(N)$
bundle of degree $m$ on which the two-dimensional gauge theory is
defined, then the corresponding flat gauge bundle over $L(p,q)$
is~\cite{Furuta} $\pi^*({\cal P})\otimes\pi^*({\cal T}^{-m})$.

Finally, we can compute the value of the Chern-Simons action
(\ref{CSaction}) on a generic classical solution on $L(p,q)$ by using
the fact that the connection (\ref{YMinstpullback}) has constant
central curvature. After taking into account the quantum shift of the
Chern-Simons level $k\to k+N$, one finds
\bea
S_{U(N)}^{\rm CS}\big(\mbf N\big):=S_{U(N)}^{\rm CS}\big(A(\mbf N)
\big)&=&\frac{\ii(k+N)}{4\pi}\,\int_{L(p,q)}\,\Tr\big(A(\mbf N)
\wedge\dd A(\mbf N)\big)\nonumber\\[4pt] &=&\frac{\ii(k+N)}{4\pi}\,
\sum_{m=0}^{p-1}\,(2\pi\,m)^2\,N_m~\int_{L(p,q)}\,\kappa\wedge
\dd\kappa \ .
\label{SCSflatvalues}\eea
By using eq.~(\ref{c1integral}) we arrive at the final form
\beq
S_{U(N)}^{\rm CS}\big(\mbf N\big)=\frac{\pi\ii(k+N)\,q}p\,
\sum_{m=0}^{p-1}\,N_m\,m^2 \ .
\label{SCSflatfinal}\eeq
This result confirms the conjectured formula~\cite{Hansen} for the set
of values of the Chern-Simons action functional of flat
$G$-connections on $L(p,q)$ in the case of gauge group $G=U(N)$. It
also agrees with the classical part of the partition function
(\ref{ZCSUNLpqm}), upon using the identification (\ref{gYMkCS}).

\subsection{Semi-Classical Expansion\label{CSLpqPartFn}}

We will now describe how to compute the one-loop fluctuation
determinants needed to write down the localization of the partition
function (\ref{CSpartfn}) onto a sum over the classical solutions
constructed in Section~\ref{FlatLpq} above. For this, we use the
well-known surgery construction of the Lens space
$L(p,q)$~\cite{Freed:1991wd,Jeffrey:1992tk}. Choose a
pair of integers $r,s$ which satisfy the Diophantine equation
$s\,q-r\,p=1$. Then the Seifert manifold $L(p,q)$ is obtained from
$\PP^1\times\sphere^1$ by removing a solid torus
$\disk^2\times\sphere^1$ (with disk $\disk^2\subset\PP^1$) and gluing
it back by twisting its torus boundary by the $SL(2,\zed)$ modular
transformation
\beq
{\sf M}=\begin{pmatrix}q~&~r\\p~&~s\end{pmatrix} \ .
\label{Mglue}\eeq
The basis element $(q,p)\in
H_1(\sphere^1\times\sphere^1,\zed)\cong\zed^2$ specifies the slope of
the meridian of the boundary torus, while $(r,s)$ gives the slope of
the longitude. With the continued fraction expansion
(\ref{pqcontfrac}), the gluing matrix (\ref{Mglue}) can be cast in the
form
\beq
{\sf M}={\sf S}~{\sf T}^{e_1}~{\sf S}~\cdots~{\sf S}~
{\sf T}^{e_\ell}~{\sf S} \ ,
\label{MglueST}\eeq
where
\beq
{\sf S}=\begin{pmatrix}0&~-1\\1&~0\end{pmatrix} \qquad \mbox{and}
\qquad {\sf T}=\begin{pmatrix}1~&~1\\0~&~1\end{pmatrix}
\label{SL2Zgens}\eeq
are the standard generators of $SL(2,\zed)$ obeying the relations
${\sf S}^2=({\sf S}~{\sf T})^3=\id$. In particular, the matrix
(\ref{InterMatrix}) in this context gives the linking matrix of the
framed surgery link with framings of components specified by the
integers $e_i$.

According to the gluing rules of topological quantum field
theory~\cite{Witten:1988hf}, the partition function (\ref{CSpartfn})
of Chern-Simons theory in the canonical two-framing of $L(p,q)$ may
thus be computed (up to normalization) as the matrix
element~\cite{Freed:1991wd,Jeffrey:1992tk}
\beq
Z_{U(N)}^{\rm CS}\big(L(p,q)\,,\,k\big)={\cal R}({\sf M})_{0,0} \ ,
\label{ZCSsurgery}\eeq
where $\cal R$ is the representation of the mapping class group on the
finite-dimensional quantum Hilbert space of Chern-Simons gauge theory
on $\PP^1\times\sphere^1$. In the Verlinde basis of level~$k$
integrable representations $R$ of the $U(N)$ WZW model, the generators
(\ref{SL2Zgens}) are represented as~\cite{Jeffrey:1992tk}
\beq
{\cal R}({\sf S})_{R,Q}=S_{R,Q} \qquad \mbox{and} \qquad
{\cal R}({\sf T})_{R,Q}=\delta_{R,Q}~T_R
\label{RSTVerlinde}\eeq
in terms of the amplitudes (\ref{Smatrixdef}) and (\ref{TRdef}) with
the identification (\ref{gYMkCS}). We can thus write
(\ref{ZCSsurgery}) as
\beq
Z_{U(N)}^{\rm CS}\big(L(p,q)\,,\,k\big)=\sum_{R_1,\dots,R_\ell}\,
S_{0,R_1}\,S_{R_1,R_2}\,\cdots\,S_{R_{\ell-1},R_\ell}\,S_{R_\ell,0}~
T_{R_1}^{e_1}\,\cdots\,T_{R_\ell}^{e_\ell} \ .
\label{ZCSLpqreps}\eeq
Although formally identical to the q-deformed gauge theory partition
function (\ref{Zpq}) with $\theta_i=0$, in (\ref{ZCSsurgery}) one
quantizes the Yang-Mills coupling as in (\ref{gYMkCS}) and restricts
the sum to {\it integrable} representations of the $U(N)$ gauge group
at level $k\in\nat_0$. Similar correspondences between q-deformed
Yang-Mills theory and Cherns-Simons theory on circle bundles have been
noted in~\cite{Blau:2006gh}.

The sums over integrable representations in (\ref{ZCSLpqreps}) can be
written in terms of weight vectors $\mbf n(R_i)$ with
$0\leq n_a(R_i)\leq N+k-1$. Using the explicit matrix elements in
(\ref{Smatrixdef}) and (\ref{TRdef}), this writes the Chern-Simons
partition function as a lattice Gauss sum. To cast (\ref{ZCSLpqreps})
as a sum over classical solutions, one uses the reciprocity formula
for Gauss sums to resum the expansion over weight vectors. This
calculation was first performed for the case of an $SU(2)$ gauge group
in~\cite{Jeffrey:1992tk}, and more recently extended in~\cite{Hansen}
to arbitrary simple Lie groups $G$. We will not enter into the
intricate details of this calculation here, which are analogous to the
Poisson resummation carried out in Section~\ref{InstExp}. Dropping
irrelevant overall normalization factors, for $G=U(N)$ one
finds~\cite{Hansen}
\beq
Z_{U(N)}^{\rm CS}\big(L(p,q)\,,\,k\big)=\sum_{\mbf m\in\zed_p^N}\,
\e^{-\frac{\pi\ii(k+N)\,q}p\,\mbf m^2}~{\cal W}_{U(N)}^{\rm fluct}(p,q;
\mbf m)
\label{ZCSLpqfinal}\eeq
where
\beq
{\cal W}_{U(N)}^{\rm fluct}(p,q;\mbf m)=\sum_{w\in S_N}\,\varepsilon(w)~
\e^{-\frac{2\pi\ii}{(k+N)\,p}\,w(\mbf\rho)\cdot\mbf\rho}~
\e^{\frac{2\pi\ii}p\,\mbf m\cdot(q\,\mbf\rho-w(\mbf\rho))} \ .
\label{WUNfluctdef}\eeq

In the first exponential factor of (\ref{ZCSLpqfinal}) we recognize
the Boltzmann weights of the classical Chern-Simons action
(\ref{SCSflatfinal}) evaluated on the set of critical points. The Weyl
group sums (\ref{WUNfluctdef}) thereby represent the one-loop quantum
fluctuation determinants about the classical solutions. This justifies
the identification made in (\ref{ZCSUNLpqm}), and also the analysis of
Section~\ref{2Dto4D}, after the reflections $(p,q)\to(-p,-q)$. This
defines an orientation-reversing isometry of the four-manifold
$X(p,q)$ under which the topologically twisted ${\cal N}=4$ Yang-Mills
theory is invariant, but under which the Chern-Simons and q-deformed
gauge theories are not. The remarkable feature of the calculation
performed in~\cite{Hansen,Jeffrey:1992tk} proceeding from
(\ref{ZCSLpqreps}) to (\ref{ZCSLpqfinal}) is that the final form
depends only on the integers $p$ and $q$ which uniquely determine the
Seifert space $L(p,q)$ up to isomorphism, and not on the continued
fraction expansion (\ref{pqcontfrac}). This is expected, since the
surgery construction of the Chern-Simons partition function
(\ref{ZCSsurgery}) is independent of the framing integers
$e_i$~\cite{Freed:1991wd}. Moreover, while the expansion
(\ref{pqcontfrac}) is not unique, any two such decompositions are
related by an $SL(2,\zed)$ transformation, and the Chern-Simons
partition function is invariant under the action of the mapping class
group. In marked contrast, the geometry of the Hirzebruch-Jung space
$X(p,q)$ depends crucially on the continued expansion of $\frac pq$
(mod $SL(2,\zed)$) and the corresponding gauge theory amplitudes
reflect this dependence.

\section{Instantons on Higher Genus Ruled Surfaces\label{HigherGenus}}

In this section we will address the problem of counting instantons on
the ruled Riemann surfaces~\cite{BPVdeV}, which can be described as
the total space of a holomorphic line bundle ${\cal O}_{\Sigma_g}(-p)$
of degree $p$ over a compact Riemann surface $\Sigma_g$ of genus
$g\geq1$. This non-toric manifold can be viewed as a non-compact
four-cycle in the local Calabi-Yau threefold which is the total space
of the holomorphic rank~$2$ vector bundle ${\cal
  O}_{\Sigma_g}(-p)\oplus{\cal O}_{\Sigma_g}(2g-2+p)$, as considered
by~\cite{Vafa:2004qa,Aganagic:2004js} for the problem of counting BPS
black hole microstates in four dimensions. In this case, the direct
instanton counting in four dimensions is a difficult problem, and the
two-dimensional gauge theory could thus provide valuable
insight. In~\cite{Aganagic:2004js} the q-deformed gauge
theory on $\Sigma_g$ is proposed to compute the relevant Euler
characteristic of the instanton moduli space.

In the case of gauge group $U(1)$, one can give a prediction for the
partition function of $\mathcal{N}=4$ gauge theory on ${\cal
  O}_{\Sigma_g}(-p)$ for any genus $g$. From the known partition
function of $U(1)$ gauge theory on $\Sigma_g$~\cite{Blau:1991mp}, one
can follow the prescription of Section~\ref{2Dto4D} to read off the
fractional instanton contribution directly as
\beq Z^{U(1)}_{\rm frac}({\Sigma_g})=\sqrt{\frac{4\pi}{g_{\rm YM}^2\,p
}}~\sum_{m\in\mathds{Z}}\,\e^{-\frac{4\pi^2}{g_{\rm YM}^2\,p }\, m^2 -
z\,m} \ . \eeq
On the other hand, the moduli space of $n$ regular instantons
of rank~$1$ is given by the Hilbert scheme ${\cal
  O}_{\Sigma_g}(-p)^{[n]}$ of $n$ points on the total space of the
bundle ${{\cal O}_{\Sigma_g}(-p)}$~\cite{nakabook}. The
generating function for the corresponding Poincar\'e polynomials is
given by
\be \sum_{n=0}^\infty\, P\big(t\,\big|\,{\cal O}_{\Sigma_g}(-p)^{[n]}
\big)~\e^{2\pi\ii n\,\tau} =
\prod_{m=1}^\infty\, \frac{\left(1+ t^{2m-1}~\e^{2\pi\ii m\,\tau}
\right)^{2g}}
{\left(1-\e^{2\pi\ii m\,\tau}\,t^{2m}\right)^2\,
\left(1-\e^{2\pi\ii m\,\tau}\,t^{2m-2}\right)} \ . \label{genera} \ee
By setting $t=1$ in (\ref{genera}) we get the contribution of regular
instantons to the four-dimensional partition function.
By assuming the factorization property between regular and fractional
instanton contributions we finally get the total partition function
(dropping irrelevant overall normalizations)
\be
Z^{U(1)}(\Sigma_g)= \prod_{m=1}^\infty\, \frac{\left(1+ \e^{2\pi\ii
      m\,\tau}\right)^{2g}}
{\left(1-\e^{2\pi\ii m\,\tau}\right)^2}~
 \sum_{n\in\mathds{Z}}\, \e^{\frac{\pi\ii\tau\,n^2}{p }-z\,n}
=\frac{\hat\eta(2\tau)^{2g}}{\hat\eta(\tau)^{2g+2}}~
\theta_3\left(\left.\mbox{$\frac\tau p$}\right|\mbox{$\frac{\ii z}
{2\pi}$}\right) \ .
\label{piropiro} \ee

Compared to the genus~$0$ cases considered in the previous sections,
the computation for higher rank gauge groups is much more involved in
this case. In particular, the $U(N)$ instanton partition function does
not trivially factorize into a $U(1)^N$ contribution. Let us
illustrate this point in the genus~$1$ case, wherein a complete
analysis of the two-dimensional Yang-Mills partition function and of
its relation with Chern-Simons theory has been recently carried out
in~\cite{Caporaso:2006kk}. Starting from these results, it is possible
repeat the procedure of Section~\ref{2Dto4D} to extract the
contributions of fractional instantons in four dimensions for
nonabelian gauge group. For example, for $U(2)$ gauge group and
vanishing $\theta$-angle the instanton expansion of the
two-dimensional Yang-Mills partition function on $\Sigma_1$ is given
by
\bea {Z}_{U(2)}^{q{\rm YM}}\big({\cal O}_{\Sigma_1}(-p)\,,\,
g_{\rm YM}^2\big)&=& \sum_{m_1,m_2\in\zed}\,
\left((-1)^{m_1+m_2}\frac{4\pi}{g_{\rm YM}^2\,
p}+\delta_{m_1,m_2}\,\frac{1}{\sqrt{2}}\,\sqrt{\frac{4\pi}{g_{\rm YM}^2\,
p}}~ \right)\nonumber \\ && \qquad\qquad\qquad\qquad
\times~\e^{-\frac{4\pi^2}{g_{\rm YM}^2\,
p}\,(m_1^2+m_2^2)}\nonumber\\ &&+\,
\sum_{m_0\in\zed}\,\frac{(-1)^{2 m_0+1}}{\sqrt{2}}\,
\sqrt{\frac{4\pi}{g_{\rm YM}^2\,p}}~\e^{-\frac{2\pi^2}
{g_{\rm YM}^2\, p}\,(2m_0+1)^2} \ , \label{su22}
\eea
where the coefficients of the exponentials can be identified with
the one-loop fluctuations in Chern-Simons gauge theory on a torus
bundle over the circle~\cite{Caporaso:2006kk}. It follows that the
$U(2)$ partition function for fractional instantons on the total space
of ${\cal O}_{\Sigma_1}(-p)$ is given by
\bea
{Z}^{U(2)}_{\rm frac}(\Sigma_1)&=& \sum_{m_1,m_2\in\zed}\,
\e^{-\frac{4\pi^2}{g_{\rm YM}^2\,
p}\,(m_1^2+m_2^2)}+\sum_{m_0\in\zed}\,
\e^{-\frac{8\pi^2}{g_{\rm YM}^2\, p}\,\left(m_0+\frac{1}{2}\right)^2}
\nonumber\\[4pt] &=& \theta_3\left(\left.\mbox{$\frac\tau p$}\right|0
\right)^2+\theta_2\left(\left.\mbox{$\frac{2\tau}{p}$}\right|0\right)
\ . \label{su23}
\eea

Due to the presence of the last term on the right-hand side of
(\ref{su23}), the partition function for $U(2)$ gauge group cannot be
written as the square of that for $U(1)$. This extra term is due to
the appearance of singular fixed points in the nonabelian localization
prescription on higher genus surfaces, arising from irreducible
connections of the two-dimensional gauge theory. Thus, in order to
provide a general formula in the nonabelian case for the class of
non-toric four-manifolds modelled on ${\cal O}_{\Sigma_g}(-p)$, a more
careful analysis is required.

\section{Conclusions\label{Conclusions}}

In this paper we have shown how instanton counting on the most general
four-dimensional toric singularities $A_{p,q}$ can be carried out by
studying the classical solutions of a suitable two-dimensional gauge
theory living on the necklace of $\mathds{P}^1$'s arising in their
minimal resolutions $X(p,q)$. We have found that the two-dimensional
gauge theory captures the contributions of instantons which are
stacked at the singularity. These instantons can be recovered from
pullback of the classical solutions of two-dimensional Yang-Mills
theory. Identical results have been obtained by a direct
four-dimensional analysis in~\cite{fcr}, where the contributions of
instantons which are free to move in the non-compact directions of
$X(p,q)$ have also been investigated. The appearance of these latter
configurations is more elusive in the two-dimensional gauge theory and
require suitable regularization. Due to the lack of an explicit
construction of their moduli space, a complete evaluation of their
contribution to the D0--D2--D4 brane partition function is not yet
available except for ALE spaces~\cite{Fucito:2004ry,fujii,fcr}, and
the $\mathcal{O}_{\mathds{P}^1}(-p)$ spaces for
$p=1$~\cite{Nakajima:2003pg,Nakajima:2003uh} and
$p=2$~\cite{Sasaki:2006vq}.

In contrast to the four-dimensional case, the two-dimensional gauge
theory description contains perturbative contributions coming from the
fluctuations of flat connections at the boundary of $X(p,q)$. As shown
in~\cite{Arsiwalla:2005jb}--\cite{Caporaso:2005ta,Caporaso:2005np}
in the case of the space $X(p,1)$, these
fluctuations are a crucial ingredient in reproducing the large $N$
factorization of q-deformed Yang-Mills theory into holomorphic and
antiholomorphic topological string amplitudes, in accordance with the
OSV conjecture~\cite{Ooguri:2004zv}. It would be interesting to better
understand the meaning of these perturbative corrections from the
perspective of counting black hole microstates and D-brane bound
states.

The two-dimensional gauge theory can also be applied to more
general non-toric manifolds such as the ruled Riemann surfaces studied
in Section~\ref{HigherGenus}, which are four-cycles of the local
Calabi-Yau threefold given by the total space of the bundle ${\cal
  O}_{\Sigma_g}(-p)\oplus{\cal O}_{\Sigma_g}(2g-2+p)$. In this case
the pertinent two-dimensional gauge theory is still exactly solvable.
 Its large~$N$ chiral expansion in the case $p=0$ has been carried out
 in~\cite{deHaro}. Some results for $U(1)$ gauge group at any genus
 $g$ and gauge group $U(2)$ at genus $g=1$ are derived in
 Section~\ref{HigherGenus}. It would be gratifying to corroborate
 these expectations with a direct evaluation in four dimensions.

\acknowledgments

We warmly thank U.~Bruzzo, F.~Fucito and J.~F.~Morales for helpful
discussions. We also thank B.~Fantechi, M.~Mari\~no, S.~Pasquetti and
R.~Poghossian for fruitful exchanges of ideas. We thank the organisor
of the informal meeting on topological strings held in Alessandria and
Torino in June~2006 which stimulated the research presented in this
paper. This work was supported in part by the EU-RTN Network Grant
MRTN-CT-2004-005104.

\appendix
\addcontentsline{toc}{section}{Appendices}

\section{Continued Fractions and the Cartan
  Matrix\label{Bappendix}}

In this appendix we collect some useful properties of the inverse of
the intersection matrix (\ref{InterMatrix}) based on the continued
fraction expansion (\ref{pqcontfrac}) of the rational number $\frac
pq$. These properties are used extensively in Section~\ref{qYMToric}
to simplify the final result of the Poisson resummation of the
q-deformed Yang-Mills partition function. Given
eq.~(\ref{pqcontfrac}), we construct two sequences of integers
$\{p_j\}$ and $\{q_j\}$ defined by the continued fraction expansions
\beq
\frac{p_{j-1}}{p_j}=[e_j,e_{j+1},\ldots,e_\R] \qquad \mbox{and}
\qquad \frac{q_{j+1}}{q_j}=[e_j,e_{j-1},\ldots,e_1]
\label{contfracseq}\eeq
for $1\leq j\leq\ell$, together with the initial conditions $q_0=0$
and $p_\ell=1$. They satisfy the Diophantine relations
\beq
q_i\, p_j-p_i\, q_j= p\, n_{ij}
\label{Dioseq}\eeq
where
\beq
n_{ij}=\left\{\begin{array}{ll}
0\ \ \  &\mathrm{for}\ i=j \ , \\
1\ \ \  &\mathrm{for}\ i=j+1 \ , \\
\mbox{num$[e_{j+1},\ldots,e_{i-1}]$}\ \ \  &\mathrm{for}\ i>j+1 \ ,
\end{array}\right.
\label{numdef}\eeq
and num stands for the numerator of the continued fraction
expansion. We have the obvious values
\beq
p_1=q \ , \quad q_1=1 \quad \mbox{and}
\quad  q_\R=q^\prime
\label{obviouspq}\eeq
where the integer $q'$ is defined by $q \,q^\prime\equiv1$~mod~$p$.

In terms of these sequences of integers, the inverse of the Cartan
matrix admits the elegant form~\cite{MartinecMoore}
\beq
\big(C^{-1}\big)_{ij}=\left \{\begin{array}{ll} \frac{1}{p}\, q_i\,
 p_j\ \ \  &\mathrm{for}\ \ \ 1\le i\le j\le\R \ , \\
\\
\frac{1}{p}\, p_i\, q_j\ \ \  &\mathrm{for}\ \ \ 1\le j\le i\le\R \ .
\end{array}\right.
\label{CartaninverseMM}\eeq
A simple application of the form (\ref{CartaninverseMM}) of the Cartan
matrix and of the relations (\ref{contfracseq})--(\ref{obviouspq})
shows that
\beq
\big(C^{-1}\big)_{1i}-q \, \big(C^{-1}\big)_{i\R}=
\mbox{$\frac{1}{p}$}\,\big(q_1\, p_i -p_1\, q_i\, p_\R\big)=
\mbox{$\frac{1}{p}$}\,\big(q_1\, p_i -p_1\, q_i \big)=-n_{1i} \ .
\eeq
One can also show~\cite{MartinecMoore} from these relations that $p$
is the determinant of the Cartan matrix $C$ while $q$ is the
determinant of the first minor of $C$. In the main text we make use of
the short-hand notation
\beq
h_i=n_{i1} \ ,
\eeq
and define the symmetric $\ell\times\ell$ matrix $G_{ij}$ by
\beq
G_{11}=p\, q \ , \quad  G_{1 s}=G_{s 1}=h_s \quad \mbox{and} \quad
G_{s_1s_2}= h_{s_1}\, q_{s_2}
\eeq
for $s,s_1,s_2=2,\ldots, \R$ and $s_1\le s_2$.

\end{document}